\documentstyle[12pt,epsf]{article}
\newcommand{\dia}{\begin{displaymath}}
\newcommand{\die}{\end{displaymath}}
\newcommand{\eqa}{\begin{equation}}
\newcommand{\beq}{\eqa}
\newcommand{\eqe}{\end{equation}}
\newcommand{\eeq}{\eqe}
\newcommand{\eqna}{\begin{eqnarray}}

\newcommand{\eqne}{\end{eqnarray}}
\newcommand{\eeqa}{\end{eqnarray}}
\newcommand{\eqnaa}{\begin{eqnarray*}}

\newcommand{\eqnae}{\end{eqnarray*}}

\newcommand{\bra}[1]{\mbox{$\langle #1 |$}}
\newcommand{\ket}[1]{\mbox{$| #1 \rangle$}}

\date{February 25, 1993}
\begin{document}
\begin{center}
  {\large\bf Stochastic Methods for Quantum Scattering}
\end{center}

\begin{center}
   STEFAN LENZ \\
   {\it Institut f\"ur theoretische Physik III \\
   Friedrich-Alexander Universit\"at Erlangen   \\
   Staudtstrasse 7 \\
   D-91058 Erlangen  \\
   Germany \\
   email: slenz@faupt31.physik.uni-erlangen.de }
\end{center}

\begin{center}
   {\bf ABSTRACT}
\end{center}
 Quantum scattering at zero energy is studied with stochastic
 methods. A path integral representation for the scattering
 cross section is developed.
 It is demonstrated that Monte Carlo simulation
 can be used to compare effective potentials which
 are frequently used in multiple scattering with the exact
 result. \\

\noindent {\bf 1. Introduction } \\

\noindent
Multiple scattering off nuclei is in general a complicated
many body problem, as target and
projectile degrees of freedom are strongly coupled.
The standard method for treating multiple scattering
problems is the construction of an effective one-body
optical model potential by
eliminating the target degrees of freedom.
Optical
potential calculations have been widely and very succesfully
used in the past \cite{Foldy}.

Despite their phenomenological success,
there are severe shortcomings of these models.
One example is the spectrum of kaonic atoms, where the shifts
of the lowest level require a repulsive real part of the
optical potential, in contrast to the results of conventional
fits \cite{Friedman}.
Another problem, which is conceptually even more severe, is the
absence of reliable methods for calculating inclusive cross sections,
for which optical potential models can not be applied at all, as they are
based on restriction of the target Hilbert space.

Because of these problems, alternative methods have to be
studied. For calculating ground state properties of a many
body system
beyond perturbative or mean field approximations, stochastic methods
are well established. Starting from a path integral expression
for the density matrix, an algorithm of
Metropolis type or Langevin simulation is used for path sampling
\cite{Negele}.
The advantage of these approaches is that they provide results, which
are in principle exact and can be used to develop better analytical
understanding of the physical system under investigation.\\

\noindent {\bf 2. Path integrals and scattering observables} \\

\noindent
We start from a Hamiltonian
\[
   H=H_{int}+\frac{p^2}{2m}+V
\]
which can be decomposed into an internal target Hamiltonian $H_{int}$,
a projectile kinetic energy and a projectile-target interaction $V$.
In case the projectile has no bound state, the ground state of the
system is the zero projectile momentum scattering wave function
$\Psi_{0,k=0}(x,q)$. In the
low temperature limit, the density matrix of the system is dominated
by this state \cite{Gelman}:
\beq
   \rho(x^\prime q , x q | \beta ) =
   \bra{ x^\prime q^\prime } e^{-\beta H}
   \ket{x q}
   \stackrel{ \beta \to \infty}{ = }
   e^{-\beta E_0} \left( \frac{2\pi m}{\beta} \right)^{3/2}
   \Psi_{0,k=0}(x^\prime , q^\prime)
   \Psi_{0,k=0}^{\dagger}(x , q )
   \label{eq:dens}
\eeq
$q$ denotes the target and $x$ the projectile degrees ao freedom.
$E_0$ is the target ground state energy.
In pure bound state problems, convergence is controled
by the energy gap between first excited state and ground state. Here,
the ground state of the system lies at the edge of a continuum. This
fact manifests itself in the $\beta^{3/2}$ factor in front of
eq.(\ref{eq:dens}). The slow convergence, as compared with
bound state problems, requires rather long times $\beta$, which
makes it necessary to choose observables and path sampling
techniques carefully.

{}From the left hand side of (\ref{eq:dens}),
a path integral expression can be derived
\cite{Feynman} in the standard way.
Path sampling methods, however, do not yield the path integral directly,
but give only paths sampled according to the normalized functional
\beq
   P[x(t),q(q)] := \frac{ \exp(-S[x(t),q(t)])}
              { \int d[x] d[q] \exp (-S[x(t),q(t)])}
   .
   \label{eq:exp}
\eeq
This difficulty can be solved by measuring the following functional:
\beq
   O[x(t),q(t)] = \exp \int dt [V(x(t),q(t)) - U(x(t))]
   \label{eq:O}
\eeq
The interaction between projectile and target is removed from the
numerator and replaced by an effective interaction $U$, which only acts
on the projectile degrees of freedom, like in conventional treatment
of multiple scattering physics. The advantage
here is, that the stochastic process can be used to test the
quality of the effective potential $U$.
The expectation value of $O$ in $P$ is in the limit
$\beta\to\infty$:
\[
  \langle O \rangle \stackrel{\beta\to\infty}{=}
   \frac{\Phi_0 (q) \Phi_0 (q^\prime )
   \psi_{k=0} (x^\prime) \psi_{k=0}^{\dagger} (x) }
   { \Psi_{0,k=0} (x^\prime q^\prime) \Psi_{0,k=0}^{\dagger} (xq) }
   ,
\]
where $\Phi_0 $ is the target ground state and $\psi_{k=0}$ is
the projectile scattering wave function to
the potential $U$. \\

\noindent {\bf 3. Example: Potential Scattering} \\

\begin{figure}
\epsfxsize=10.0cm
\centerline{\epsffile{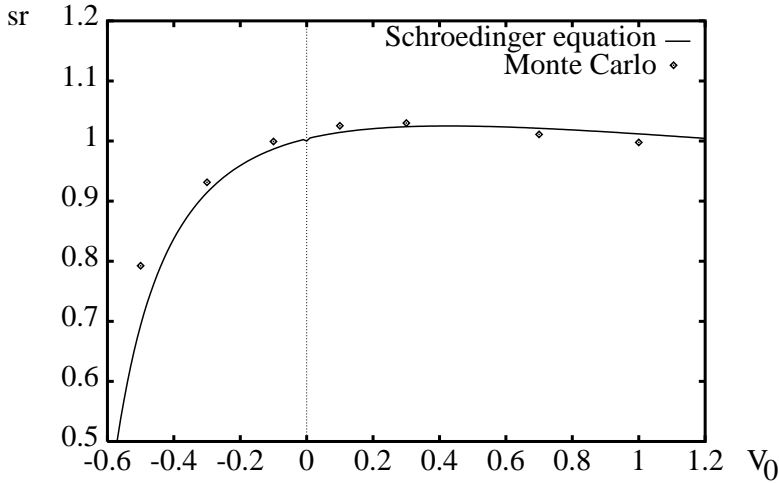}}
\caption{Ratio of cross sections of square well potential and
Gaussian potential $sr$ as function of potential strength $V_0$.
The line is the exact result, the data points are a stochastic calculation
at $\beta=100$. The path was subdivided into $200$ time intervals
$\Delta\beta=0.5$. The projectile mass was set to $m=1$.
}
\label{Abb1}
\end{figure}

\noindent
To demonstrate the feasibility of this type of calculations
I discuss potential scattering.
In this case numerical integration of the Schroedinger equation
provides exact results.
Scattering off a Gaussian potential $V_g$ is considered. As
reference potential a square well $V_w$ is used:
\[
   V_g (r) = V_0 \exp(-\frac{1}{2} r^2 ) \quad\quad
   V_w (r) = \left\{
   \begin{array}{cc}
       3\sqrt{\frac{\pi}{250}} V_0 & , r<\sqrt{5} \\
            0                      & , r>\sqrt{5}
   \end{array} \right.
\]
$V_w$ plays the role of the effective potential $U$ in
(\ref{eq:O}).
The parameters of the square well are fitted to reproduce
the first two nonvanishing moments of the Gaussian potential.
Fig.\ref{Abb1} shows the ratio of the cross sections of the
two potentials $sr:=\sigma_{V_w} / \sigma_{V_g}$
as a function of $V_0$. The line is the exact result,
the data points are obtained from a stochastic calculation
at $\beta=100$.
Path sampling was performed with a simple forward Euler scheme
Langevin algorithm \cite{Namiki}. $2.5\times 10^5$ paths were used to
measure $O$ after an equilibration run of $2.5\times 10^4$
updates.

In the range of $V_0$ where the reference potential
is already a good guess,
the stochastic calculation reproduces the exact result
within $2\%$.
Where this is no longer true, results become worse.
For $V_0=-0.5$ the stochastic result deviates about $17\%$.
Note that at this value of $V_0$ the cross section
is already $\sigma=274$, because
$V_g$ has the first bound state at $V_0=-0.66$.
There are two ways to improve the results in this region:
One possibility is to increase $\beta$.
This would require much longer calculation times, as the
autocorrelation time of the Langevin algorithm increases
like $\beta^2$.
The other possibility is to improve the reference potential.
By adjusting the parameters of the potential successively,
one can obtain results for the stochastic calculation
which do not deviate more than a few percent from the exact result,
although the same simulation parameters as for the first
calculation were used. \\

\noindent {\bf 4. Discussion } \\

\noindent
A new method for calculating elastic cross sections at
zero projectile momentum was presented. The crucial point
is that this method relies on the comparison of the full
problem with a reference problem.
This makes it possible to study questions related to the
construction of effective potentials in nuclear multiple
scattering by computer simulations, which seems to be the
natural way for a nonperturbative treatment of many body
problems.  The method can be extended to nonzero
momentum by exploiting information
contained in the $\beta$
dependence of observables. Work in this direction, as well as
multiple scattering calculations, are currently in progress.

A severe shortcoming of this method is that at this stage it
is not possible to calculate inelastic  or
inclusive cross sections. Developement of stochastic methods
for these problems seems to be promising, as scattering observables
will not depend strongly on individual nuclear states due to
summation over final states. The simple structure of experimental
data like e.g. energy loss spectra \cite{Foldy} strongly
supports this conjecture.
\\

\noindent { \bf Acknowledgement } \\

\noindent
 I would like to thank Prof.F.Lenz for contributing valuable
 ideas to this work, H.Mall for many discussions and H.Grie{\ss}hammer
 for critical reading of the manuscript.
 This work has been supported by the DFG
 Graduiertenkolleg 'Starke Wechselwirkung' Erlangen-Regensburg.
\\


\begin{thebibliography}{99}
\bibitem{Foldy}
   C.B.Dover, {\it Antinucleon-Nucleus Interactions} in: Proceedings
   of the Fourth LEAR Workshop 1987, Harwood Academic Publishers,
   Chur, 1988, page 649ff.
\bibitem{Friedman}
   E.Friedman, A.Gal and C.J.Batty, Phys. Lett. {\bf B 308} 6 (1993)
\bibitem{Negele}
   J.W.Negele and H.Orland, {\it Quantum Many-Particle Systems},
   Addison Wesley, Reading, Mass., 1988
\bibitem{Gelman}
  D.Gelman and L.Spruch, J.Math.Phys. {\bf 10} 2240 (1969)
\bibitem{Feynman}
  R.P.Feynman and A.R.Hibbs, {\it Quantum Mechanics and Path Integrals},
  McGraw-Hill, New York, 1965
\bibitem{Namiki}
  M.Namiki, {\it Stochastic Quantization},
  Springer Verlag, Berlin, Heidelberg, 1992
\end{thebibliography}
\end{document}